# Reconfiguring Data Infrastructure Ecosystem in Africa: A Primer Toward Digital Sovereignty

*Olusesan Michael Awoleye (Ph.D)*
*African Institute for Science Policy and Innovation, Obafemi Awolowo University, Nigeria*
E-mail: mawoleye@oauife.edu.ng   Phone: +234-7069197823

*The growth of the Internet and its associated technologies; including digital services have tremendously impacted our society. However, scholars have noted a trend in data flow and collection; and have alleged mass surveillance and digital supremacy. To this end therefore, nations of the world such as Russia, China, Germany, Canada, France and Brazil among others have taken steps toward changing the narrative. The question now is, should Africans join these giants in this school of thought on digital sovereignty or fold their hands to remain on the other side of the divide? This question among others are the main reasons that provoked the thoughts of putting this paper together. This is with a view to demystifying the strategies to reconfigure data infrastructure in the context of Africa. It also highlights the benefits of digital technologies and its propensity to foster all round development in the continent as it relates to economic face-lift, employment creation, national security, among others. There is therefore a need for African nations to design appropriate blueprint to ensure security of her digital infrastructure and the flow of data within her cyber space. In addition, a roadmap in the immediate, short- and long-term in accordance with the framework of African developmental goals should be put in place to guide the implementation.*

## 1.      Introduction

The bulk of the economic power of African nations have been largely traced to their dependence on natural resources which include but not limited to oil and gas in Nigeria, Libya and Kenya. As well as diamonds in democratic Republic of Congo, Botswana, South-Africa and Angola (Baumüller, Donnelly, Alex, & Markus, 2011; Idowu & Lambo, 2018). This include Gold in Ghana, Tanzania and Benin among others (Andrzej Polus & Wojciech Tycholiz, 2019; Siakwah, 2017). The new oil now has been attributed to data as an essential resource that powers the information economy as a similitude of oil that fueled the industrial economy (Hirsch, 2015). Data has also been opined as the next phase of digital transformation (Capgemini Research Institute, 2020), much more in this era of the fourth industrial revolution which requires integration of digitization with physical production tools (Naudé, 2018). The process of keeping data with utmost privacy and security is very important (Awoleye, Ojuloge, & Ilori, 2014; Thomas & Grimes, 2020). This also include the process of generating and using the same either as individuals, corporate body or at national level. The advent of Internet technology as well as the emerging technologies such as artificial intelligence, IoT, 3D printing, cloud services, among others; are largely data-driven (Bibri & Krogstie, 2020; Tao et al., 2019). With all of the opportunities which digital technology has brought to us, it has not come without its associated threats.

## 2. Digital Space in Africa and the challenge of Data Privacy and Security

The research was born out of the threats that the existing management of global digital infrastructure pose to our societies, most importantly in the Africa region. One of which among others are the challenges of data privacy and security, which could have negative effect on businesses and national security, if left unattended. Africans have been leveraging on critical resources to drive their services; such as the communication satellites providing Internet services. This include but not limited to banking services, electronic transactions, e-government services, e-mail services etc. This does not exclude cloud services such as: infrastructure as a service, platform as a service and software as a service (Dogo, Salami, & Salman, 2013; Ilyas, Ahmad, & Saleem, 2020; Kwet, 2019). The question about what happens to our data before they are delivered to the final destination is unclear; since data are routed through root servers in the United State (Nocetti, 2015). Other question that may come to mind among others is, how much control do Africans have over the technologies adopted? All of these bothers on the perceived technological and governance hegemony which has been levied against the United State as it relates to Internet sovereignty (Budnitsky & Jia, 2018; Pohle & Thiel, 2020). In addition, another author also reported an agenda termed "catch-all term" which was reportedly related to a disclosure of an intelligence of the United States to dominate the global digital space, which does not exclude the countries of Africa (Couture & Toupin, 2019). All of these are pointing to the fact that our national security intelligence and financial data may be susceptible to manipulation and control by others, which is not desirable (Bozhkov, 2020). This agenda however may have been stealthily carried out following reports as noted in extant literature. The situation in Nigeria for example suggest that the digital ecosystem is somewhat porous, much more as it relates to electronic services (Awoleye, Ojuloge, & Ilori, 2014; Bakare, 2015; Salimon, Yusoff, Sanuri, & Mokhtar, 2015). For instance, majority of the financial houses depend solely on foreign proprietary software applications for their daily financial transactions. For example, it was reported that an Indian based firm developed Finacle, which is the most prevalent software used among the commercial banks in Nigeria (Awoleye, 2015). In the same vein, the websites of those banks which are the platforms for their online banking services were noted to be outsourced and are mostly developed and managed by foreign developers. This include hosting of the same by International firms, outside the shore of Nigeria (Awoleye, Okogun, & Siyanbola, 2013). The source code however for these software are not available to the banks that have acquired them. Further to this, China as a country reportedly 'gifted' the African Union a headquarters building which was later discovered to have been bugged to eavesdrop state secrets. China was noted to have built the Internet network architecture and allegedly inserted a backdoor server that allowed it to transfer data stealthily to Shanghai[1] (Meservey, 2020). This however has been speculated to be the situation for other 186 buildings globally (40 of which are in Africa); which China has either built or renovated by her tech giant -Huawei (Meservey, 2020). This has succeeded for a number of years before it was discovered, imagine what may have been lost. How do we quantify the extent of damage this may have done to the continent? In addition, it was reported by Kwet (2019) that the United State in partnership with corporations in South African tapped the Internet backbone, collected trillions of telephone calls, emails as well as banking and social networking data.

---

[1] https://qz.com/africa/1192493/china-spied-on-african-union-headquarters-for-five-years/

## 3. Cases of Digital Hegemony in Africa

It has been reported in Literature that the nations in the Global South are more vulnerable to the digital supremacy agenda, the precursory activities are evident in many countries in Africa (Olivier, 2018; Pinto, 2018). Some of which are contrary to the general data protection regulation of the European union and as applicable to Africa (Thomas & Grimes, 2020; Wolford, 2020). For example, in South Africa, Nigeria and Kenya; tech giants such as Google, Facebook and Instagram dominate the online advertising space, to the detriment of the local media outfits (Kwet, 2019). In addition, Uber was reported to have disrupted the road transport businesses, which was reiterated to have affected a great proportion of the indigenous transport individuals and company's business to a large extent (Damle, 2018; Jenk, 2015). The internet thus remains the platform where some of these services thrive, without which ecommerce activities (Tamilarasi & Elamathi, 2020), Uber services (Tong, Zhou, Zeng, Chen, & Shahabi, 2020; Werbach, 2015), online advertisement (Bruntha, Yasmeen, Indirapriyadharshini, & Giri, 2019; Ching, Tong, Chen, & Chen, 2013; Sanne & Wiese, 2018) among others could not have succeeded. The current infrastructural design has provided imperial control on a number of digital services (Couture & Toupin, 2019; Nocetti, 2015; Pohle & Thiel, 2020). In the first instance, the western giants have put access to publications behind paywalls. In the same vein, Spotify, Netflix among others are global tech giants, which have taken over and colonised the entertainment industry globally (Kwet, 2019). They have provided services such as online streaming of movies and music for meagre fees. This has in no small way perturbed the traditional process of doing the same, thereby disrupting the value chain to eliminate the indigenous micro-business that were previously involved in the distribution (Kwet, 2019). It is worth noting to state that militating against this control of digital space should not be negotiable. Kwet (2019) submitted that software corporations such as Microsoft; market proprietary software do not supply the source codes along with the applications. By this, the users cannot change the behavior of the software, which thus gives the corporation absolute ownership to the so-called proprietary licensing. This is however viewed as a form of digital colonialism (Kwet, 2019; Pinto, 2018; Pohle & Thiel, 2020). It is believed that the ability to control the flow of information and management of the same has the propensity to go a long way in bringing political stability which is a precursor to national economic development (Bozhkov, 2020). For example, digital technologies was noted to have facilitated the political instability that erupted during the Arab spring (Nocetti, 2015). It is in this context that China put forward five main priorities as strategies to take control of their cyber space (Bozhkov, 2020; Hong & Goodnight, 2020). In the same vein, Russia as a state took pragmatic steps toward retracting their data infrastructure. Of note is their request to the United State to relocate all servers to Russia on which Russian citizens' personal data were stored. Russia also requested the hosting of their country code top level domain ".ru" to be hosted in Russia. (Budnitsky & Jia, 2018; Nocetti, 2015).

## 4. Some Digital Initiatives and Infrastructural Development in Africa

Much of what we know about the connectivity of critical digital infrastructure in Africa and its impacts shows that African nations have invested grossly to connect the continent to the global Internet backbone (Friederici, Ojanperä, & Graham, 2017; MTN Group sustainability report, 2019). MTN as the leading telecommunication service provider in Africa has provided 22,204 skilled employments and has also invested over R118 billion on capital projects (MTN Group sustainability report, 2019). Of note also is GLOBACOM (an indigenous service provider in Nigeria) invested a sum of $800 million to build high-capacity fibre-optic cable known as Glo-1. A

submarine cable which connects the United Kingdom to Nigeria. Also in North Africa, the Egyptian government started a Smart Village Cairo Initiative in 2001 that was a public private partnership initiative tech park for software development. As an offshoot of this, the growth of the digital ecosystem in Egypt was reportedly attracted USD 59million in 2018 alone among other initiatives. Representing the Eastern Africa, Kenya Safaricom in 2007 introduced M-Pesa which is one of the first mobile payment system which originated in Africa and has been widely accepted (Naudé, 2018). This revolution has enabled the creation of other related products that ride on the technology in Kenya economy and beyond. Some of which among others are: the M-KOPA, Powerhive and ecommerce innovations (e.g. Twiga foods, Copia global etc). On efforts towards data infrastructure, the Kenyan government was noted to have invested in four undersea fibre-optic cables in 2009 to improve on the quality of Internet connectivity as well as to reduce the costs of access. These efforts are somewhat spontaneous as most African nations lack adequate policy frameworks to drive digital projects in their domains (Friederici et al., 2017). The existing policies were noted to be deficient and somewhat outdated.

5. **Policy suggestions**

The following are the highlights of some necessary policy directions that has the propensity to put Africa nations in her rightful place as it relates to the control of her data infrastructure.

- There is a need to create an ***encryption/decryption gateway*** (engine) using appropriate algorithm; to transform documents going-out or coming-in to the cyberspace of Africa to/from cloud services.
- Additional ***internet exchange points*** may be necessary to enable service providers interconnect to terrestrial fibre backbones in order to facilitate low bandwidth cost. Including building globally competitive ***Data Centres*** to enable continental hosting of websites, files, email among other services.
- There should be continental ***Communication satellites***, which will have its footprints across cooperating Africa countries. In this perspective, the African Union has a great role to play as this may require a policy for its governance and sustainability.
- The linkages of the academia to the Industry has been week in most of the African nations. To eliminate the traditional challenges inhibiting the required linkages, ***a digital platform (model) for patent listings and general research outputs*** could be created. In the same vein, as part of the input that should be transmitted to the industries who has the capacity to commercialise innovations are the research outputs from the knowledge institutions. This model should also ***digitalize and curate*** all research outputs including ***student's projects and researches***. To this end, projects/theses from 1990 to 2020 could be considered to populate the digital database.
- To provide access to the knowledge base, an African version of ***Federal Identity management system*** could be adapted and developed. This will give access to relevant institutions including the industries who may have registered with the designated African federation to provide access to the knowledge base.
- Adequate and appropriate ICT policy frameworks to drive digital projects should be developed and where they exist; it should be reviewed to accommodate the new thinking.